\documentclass[aps,prb,twocolumn,showpacs,floatfix,superscriptaddress,
citeautoscript
,cite]{revtex4}

\usepackage{graphicx} 
\usepackage[utf8]{inputenc}
\usepackage[T1]{fontenc}
\usepackage{lmodern}
\usepackage{color}
\usepackage{multirow}
\usepackage{amsmath}
\usepackage{amssymb}
\usepackage{rotating}

\begin{document}

\title{Anomalous Raman Spectra and Thickness Dependent Electronic properties of
WSe$_{2}$}

\author{H. Sahin}
\email{hasan.sahin@ua.ac.be}
\affiliation{Department of Physics, University of Antwerp, Groenenborgerlaan
171, B-2020 Antwerpen, Belgium}

\author{S. Tongay}
\email{tongay@berkeley.edu}
\affiliation{Department of Materials Science and Engineering, University of
California, Berkeley, California 94720, United States}

\author{S. Horzum}
\affiliation{Department of Physics, University of Antwerp, Groenenborgerlaan
171, B-2020 Antwerpen, Belgium}
\affiliation{Department of Engineering Physics, Faculty of Engineering, Ankara
University, 06100 Ankara, Turkey}

\author{W. Fan}
\affiliation{Department of Materials Science and Engineering, University of
California, Berkeley, California 94720, United States}

\author{J. Zhou}
\affiliation{Department of Materials Science and Engineering, University of
California, Berkeley, California 94720, United States}

\author{J. Li}
\affiliation{Institute of Semiconductors, Chinese Academy of Sciences, PO Box
912, Beijing 100083, People's Republic of China }

\author{J. Wu}
\affiliation{Department of Materials Science and Engineering, University of
California, Berkeley, California 94720, United States}
\affiliation{Materials Sciences Division, Lawrence Berkeley National Laboratory,
Berkeley, California 94720, United States}

\author{F. M. Peeters}
\affiliation{Department of Physics, University of Antwerp, Groenenborgerlaan
171, B-2020 Antwerpen, Belgium}

\date{\today}


\begin{abstract}

Typical Raman spectra of transition metal dichalcogenides (TMDs) display two
prominent peaks, $E_{2g}$ and $A_{1g}$, that are well separated from each other.
We find that these modes are degenerate in bulk WSe$_2$ yielding one single
Raman peak. As the dimensionality is lowered, the
observed peak splits in two as a result of broken degeneracy. In contrast to
our experimental findings, our phonon dispersion calculations reveal that these
modes remain degenerate independent of the number of layers. Interestingly,
for minuscule biaxial strain the degeneracy is preserved but once the
crystal symmetry is broken by uniaxial strain, the degeneracy is lifted.
Our calculated phonon dispersion for uniaxially strained WSe$_2$ shows a perfect
match to the measured Raman spectrum which suggests that uniaxial strain exists
in WSe$_2$ flakes possibly induced during the sample preparation and/or as a
result of interaction between WSe$_2$ and the substrate. Furthermore, we find
that WSe$_2$ undergoes an indirect to direct bandgap transition from bulk
to monolayers which is ubiquitous for semiconducting TMDs. These results not
only allow us to understand the vibrational properties of WSe$_2$ but also
provides detailed insight to their physical properties.

\end{abstract}

\maketitle

\section{Introduction}

Owing to its extraordinary properties~\cite{novo, geim}, graphene has already
been implemented in various kinds of applications/devices~\cite{miao, trans}
and led to the emergence of a new class of materials; ultra-thin two-dimensional
crystal structures. Nowadays, among the members of this new era, especially the
ultra-thin transition metal dichalcogenides (TMDs) have attracted considerable
interest.\cite{science 2011, prb-2002, nature-nano,apl-sefa} Even though, they
are only few-atom-thick, MX$_{2}$-type structures have remarkable chemical and
mechanical stability\cite{science 2011,nature-nano, mx2} and offer new physics
as various quantum confinement effects amplified in quasi-two dimension
\cite{PRL-105,hartwin,prb84,apl99}. As a result of this
confinement effect, the band gap increases and transforms to a direct band gap
with decreasing number of layers which makes them promising candidates for
nanoscale field-effect transistors and for solar cell
applications.\cite{sefamose2, jkang} Recently, possibility of vacancy creation
in TMDs under electron irradiation,\cite{hannu} bandgap transition in tungsten
dichalcogenides\cite{zhao}, existence of tightly bound negative
trions\cite{mak} and strain-engineered electronic properties have been
reported.\cite{feng} Furthermore, we reported that with strain
application MoSe$_{2}$, and possibly other TMDs, show significant red shift in
their Raman spectrum and undergo a direct to indirect bandgap
transition.\cite{SeydaMoSe2}

Synthesis and characterization of tungsten diselenide (WSe$_{2}$) has been an
active field of research with applications in photovoltaic and photoconductive
devices and recently monolayer WSe$_{2}$ has become a popular choice for
nanoscale devices\cite{hui}. Here, we present an experimental and theoretical
investigation of the electronic properties and lattice dynamics of bulk, few
layer and single layer WSe$_{2}$. We find that the $A_{1g}$ and $E_{2g}$ modes
are almost degenerate for bulk WSe$_2$ whereas these modes are well separated
for other members (MoS$_{2}$, MoSe$_{2}$ and WS$_{2}$) of the TMDs.
Interestingly, this degeneracy is lifted as the dimensionality is lowered
from 3D (bulk limit) to 2D (monolayer) where the $A_{1g}$ and $E_{2g}$ modes are
separated by $\sim$12 cm$^{-1}$. Calculated vibrational spectrum show that the
lifting the degeneracy is closely related to the uniaxial strain induced on
monolayer WSe$_2$ due to interaction with the substrate and/or sample
preparation procedure. Lastly, we show that the band gap of WSe$_2$ goes through
rather 'soft' indirect to direct band gap crossover from bulk to monolayer and
the band gap shows almost triple band degeneracy for bi and tri-layers as
evidenced by our photoluminescence measurements.

\section{Experimental details and Computational methodology}

Monolayer and  few-layer WSe$_2$ flakes were exfoliated from WSe$_2$ single
crystals (2Dsemiconductors.com) onto 90nm SiO$_2$/Si (MTI Inc.) substrates using
conventional mechanical exfoliation technique. The thickness of the WSe$_2$
flakes was confirmed by three complementary methods, atomic
force microscopy (AFM), Raman spectroscopy, and photoluminescence (PL).
Non-contact AFM line scans on the monolayer WSe$_2$, typically resulted in
$\sim$0.7 nm step height corresponding to single unit cell lattice constant for
WSe$_2$~Fig.\ref{fig1}(d). PL and Raman measurements were performed using very low power intensity
(10$\mu$W/$\mu$m$^2$) on 2-3$\mu$m$^2$ circle to avoid local heating or damaging
effect. The results presented in this manuscript were reproduced
on more than fifty samples. 

Theoretical calculations for equilibrium and
strained structures were carried out in the framework of density functional
theory (DFT), using the projector augmented wave (PAW) method\cite{paw} as
implemented in the VASP code.\cite{kresse} The generalized gradient
approximation (GGA) of Perdew-Burke-Ernzerhof (PBE) was used for the
exchange-correlation functional \cite{pbe}. To calculate the Raman spectrum of
single layer WSe$_2$ under biaxial (uniaxial) strain, hexagonal (rectangular)
unitcell with one side parallel to the direction of stretch is taken into
account. Phonon frequencies and phonon eigenvectors were calculated by using
both the density functional perturbation theory (DFPT)\cite{dfpt} and the Small
Displacement Method.\cite{alfe}

\section{RESULTS and DISCUSSIONS}

\subsection{Thickness-Dependent Electronic Properties}

\begin{figure}
\includegraphics[width=8.5cm]{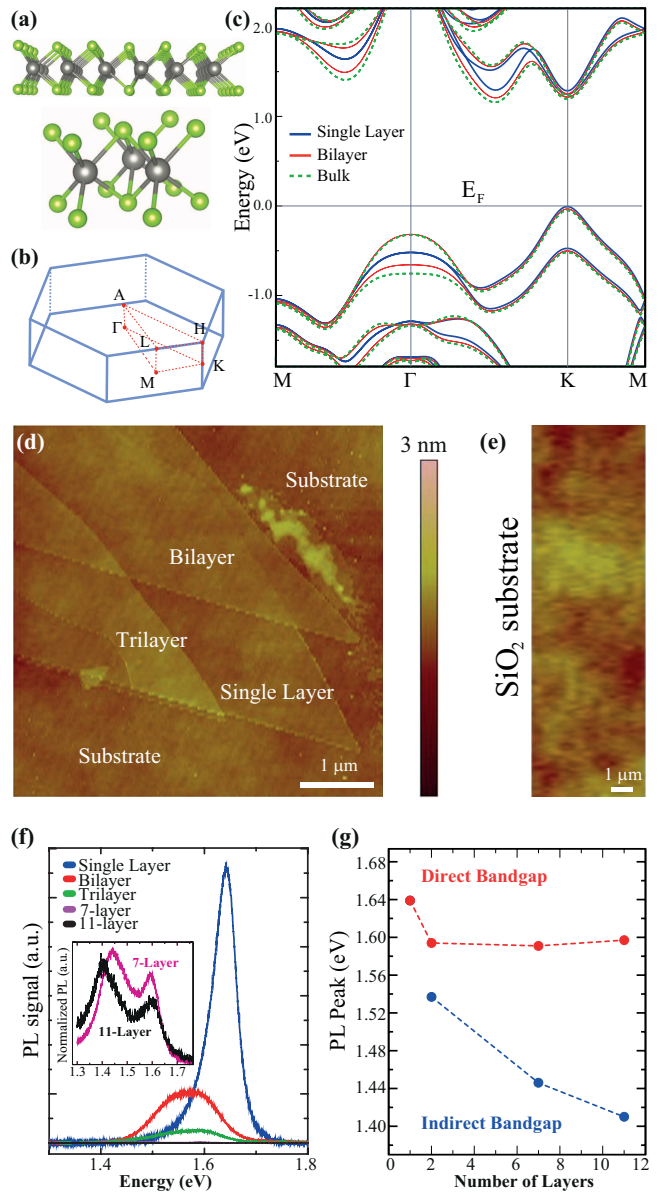}
\caption{\label{fig1}
(Color online) (a) Side and tilted view of single layer WSe$_{2}$. (b) Brillouin
Zone of WSe$_{2}$ and (c) GGA+SO band structures. Fermi level is set to zero.
Blue, red and green colors are for single layer, bilayer and bulk structures,
respectively. (d) Atomic force microscopy (AFM) images taken on WSe$_2$/SiO$_2$
and (e) bare SiO$_2$ substrate. AFM scan on sapphire substrate is given as
comparison. (f) PL measurements on WSe$_2$ flakes with various thicknesses and
(g) peak positions.}
\end{figure}

Similar to graphite, WSe$_{2}$ crystals possess lamellar structure with Bernal
stacking where the individual layers are weakly coupled to the adjacent layers
by van der Waals (vdW) forces. As shown in Fig. \ref{fig1}(a-b), each monolayer
WSe$_{2}$ (1H-WSe$_2$) consists of Se-W-Se atoms wherein tungsten atoms are
sandwiched between trigonally arranged sheets of selenium atoms. Our
calculations reveal that bulk WSe$_{2}$ has an 1.21 eV indirect band gap where
the valence band maximum (VBM) is located both at the K symmetry point while the
conduction band minimum (CBM) is along the $\Gamma$-K direction (Fig.
\ref{fig1}(c) green-dashed curve). We note that conduction bands located at K
and $\Gamma$-K is only separated by 0.04 eV in energy and hence we expect that
its electronic properties will be strain and dimensionality dependent as a
result of significant changes in the hybridization. Since bulk WSe$_2$ is an
indirect band gap semiconductor, the photoluminescence signal is expected to be
rather weak for bulk WSe$_{2}$ as observed in Fig. \ref{fig1}(f).
Interestingly, 7 to 11 layers display two distinct PL peaks  (hot luminescence)
located at 1.39 eV and 1.59 eV where the former probes the indirect band gap
($\Gamma$ to $\Gamma$-K) and the latter is associated with the direct band gap
transition (K to K) (see inset of Fig. \ref{fig1}(f)). We note that the overall
PL signal measured on few-layer flakes is orders of magnitude weaker in
intensity as compared to bilayer and monolayer WSe$_2$.

Electronic band structure calculations for bilayer and monolayer WSe$_2$ show
that the band gap increases from 1.21 eV in bulk to 1.23 eV (direct-indirect
almost degenerate) for bilayer and 1.25 eV (direct) for monolayer WSe$_2$
Fig.~\ref{fig1}(c). Experimentally observed PL signal for bulk, bilayer and
monolayer WSe$_2$ peaks at 1.57, 1.57 and 1.64 eV, respectively
(Fig.\ref{fig1}(f)). We see that in contrast to the general case, for WSe$_2$,
LDA and GGA band dispersions differ significantly. While GGA finds single layer
WSe$_2$ as a semiconductor with a direct bandgap of 1.56 eV, LDA gives indirect
bandgap of 1.68 eV. Here, both LDA and GGA finds VBM at the K symmetry point and
CBM is located at K (in between $\Gamma$ and K) point in GGA (LDA). It appears
that the GGA exchange correlation better approximates the ground state
characteristics of the WSe$_2$ crystal structure.  

Here it is also worth to note that
spin-orbit interaction results in two significant effects on the electronic band
dispersion: (i) shifting down the valence band energies at the $\Gamma$ point
(ii) band splitting at the vicinity of K and M symmetry points. Clearly,
the existence of an intrinsic electric field breaks
the inversion symmetry in the crystal structure. Therefore, it is seen that the
degeneracy of the doubly degenerate valence and conduction bands of single
layer WSe$_2$ are removed by spin-orbit interaction and a band splitting occurs.
Furthermore, when the spin-orbit interaction is taken into account in our calculations, conduction
band edges at K and K-$\Gamma$ points show band splitting that allows various
direct and indirect transitions (even in few-layered WSe$_2$ structures). In our
study, all the band structure calculations presented are performed by
considering spin-orbit interaction together with GGA.

Previously, indirect to direct band crossover has been observed in other
transition metal dichalcogenides~\cite{PRL-105,sefamose2,SeydaMoSe2} and is
consistent with our results. However, we note that the indirect to direct
transition is rather steep for MoS$_2$ where it's bilayer form is an indirect
band gap (1.6 eV) semiconductor while monolayer structure has a direct band gap
(1.9 eV). In contrast, bilayer WSe$_2$ possesses almost triple band gap
degeneracy where the K$\rightarrow$K and
K$\rightarrow$$\Gamma$-K gap values are almost degenerate and the
difference between first and second CBMs is $\Delta^{CBM}$(K)=40 meV which is
close to thermal broadening ($\sim$30meV). As a result of this band degeneracy,
the integrated PL intensity (Fig.~\ref{fig1}(e)) of bilayer WSe$_2$ is of the
same order of magnitude as in monolayers, i.e I$_{1L}$/I$_{2L}$$\sim$~1-10,
which compares with 100-1000 for MoS$_2$. We also note that the PL signal for
bilayer WSe$_2$ is rather broad and can be described by at least two Lorentzian
peaks. As the number of layers increases, the band degeneracy is gradually
lifted, WSe$_2$ becomes truly an indirect band gap semiconductor and the PL
intensity decreases by orders of magnitude.

\begin{table}
\caption{Calculated direct and indirect transitions between VBM and
CBMs. The difference between the first and second CBM at K point is also given
as $\Delta$$^{CBM}$(K). Experimental value is given in parenthesis. All energies
are given in eV.}
\label{table}
\begin{center}
\begin{tabular}{cccccccccccccccc}
\hline  \hline
WSe$_{2}$ &\underline{$\Gamma$ $\rightarrow$ $\Gamma$-K} &
\underline{$\Gamma$ $\rightarrow$ K} & \underline{K $\rightarrow$ K} &
\underline{K $\rightarrow$ $\Gamma$-K} & $\Delta^{CBM}$(K)  \\

Bulk     & 1.48 & 1.52 & 1.25 & 1.21 & 0.04    \\

Bilayer  & 1.51 & 1.52 & 1.24 & 1.23 & 0.04   \\

1-Layer  & 1.83 & 1.78 & 1.25(1.64) & 1.30 & 0.04    \\

\hline \hline
\end{tabular}
\end{center}
\end{table}

\subsection{Anomalous Lattice Vibrations: Breaking the degeneracy of A$_{1g}$ and E$_{2g}$ modes at reduced dimensions}

\begin{figure}
\includegraphics[width=8.5cm]{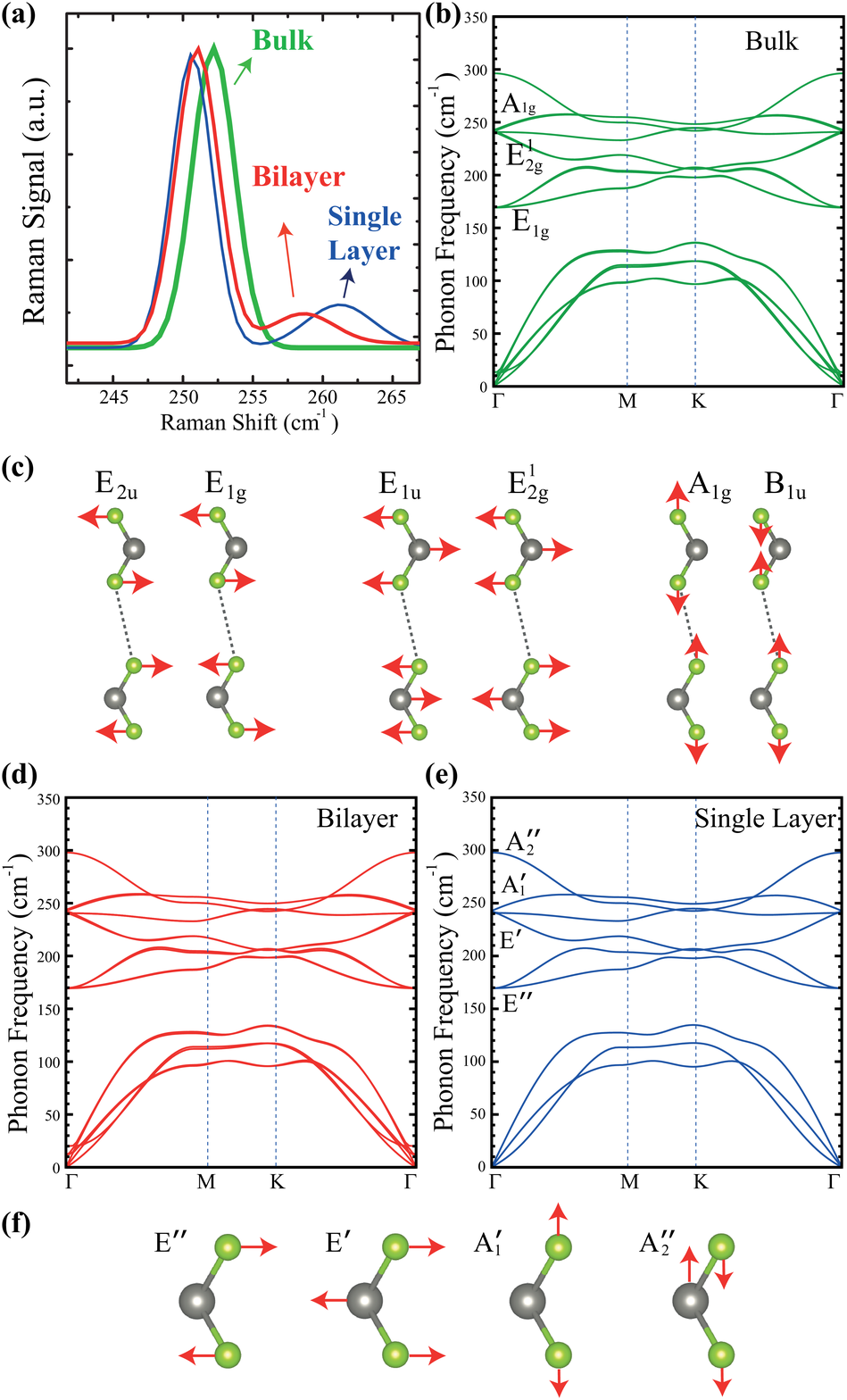}
\caption{\label{fig2}
(Color online) (a) Raman shift for 1-layer, 2-layer and bulk WSe$_{2}$.
(b) Phonon spectrum of bulk WSe$_{2}$ and (c) atomic displacements for optical
modes between 100-250 cm$^{-1}$. (d) Phonon dispersions for bilayer and (e)
single layer structure and (f) atomic displacements for optical phonon modes of
single layer WSe$_{2}$.}
\end{figure}

\begin{figure*}
\includegraphics[width=18cm]{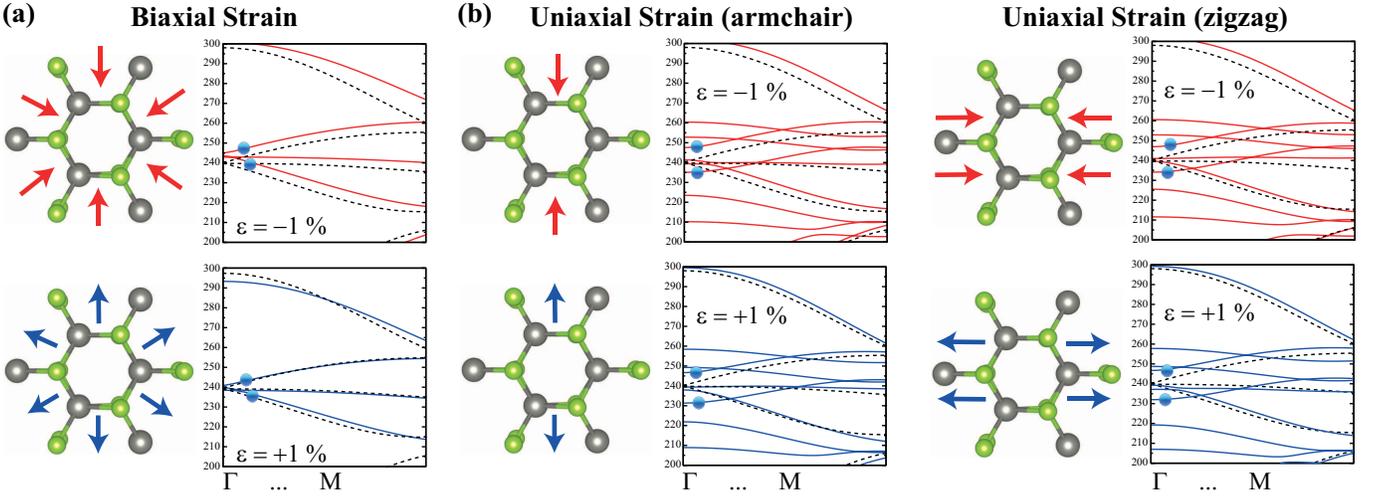}
\caption{\label{fig3}
(Color online) Application of biaxial, uniaxial-armchair and uniaxial-zigzag
strain to monolayer WSe$_2$. Compressive and tensile strain are shown by
inward and outward arrows, respectively. Phonon dispersion of unstrained 
WSe$_2$ is shown by dashed line. Raman-active branches are labeled by
blue dots.}
\end{figure*}


Next we turn our attention to the anomalous lattice vibrations of WSe$_2$.
For determination of lattice dynamics we use Raman spectroscopy which is one of
the most useful non-destructive technique for the characterization of
low-dimensional materials. Our measurements reveal that the
most prominent Raman peak for bulk WSe$_2$ is located at 252.2 cm$^{-1}$
(Fig.~\ref{fig2}(a)) while other semiconducting TMDs are characterized by well
separated  $E_{g}$ and $A_{g}$ Raman peaks. The stark difference
between WSe$_2$ and other TMDs already points towards an anomaly in the phonon
dispersion of WSe$_2$. Following from Fig.~\ref{fig2}(b), calculated phonon
dispersion confirms that bulk WSe$_2$ displays only one strong vibrational mode
around 250 cm$^{-1}$ consistent with our experiments. Interestingly, for flakes
thinner than four layers, an additional peak appears at roughly 5-11 cm$^{-1}$
above the first order Raman peak. We find that the frequency difference between
these two peaks is 5, 6 and 11 cm$^{-1}$ for trilayer, bilayer, and monolayer
WSe$_2$ flakes, respectively. This peak is absent in few-layer
flakes does not correspond to any new/additional Raman modes in the
calculated phonon dispersion curve, which describes the Raman spectrum well.

Bulk and single layer WSe$_2$ can be classified in the space group P63/mmc and
P6m2, respectively. Similar to the MoS$_{2}$ and MoSe$_{2}$
counterparts,\cite{SeydaMoSe2} the unitcell of bulk WSe$_2$ has eighteen phonon
branches corresponding to six acoustic and twelve
optical phonon modes. While the modes at $\sim$170 ($\sim$300) cm$^{-1}$ are
4-fold (2-fold) degenerate, six modes become almost degenerate at $\sim$250
cm$^{-1}$ at the  $\Gamma$ point. For a better understanding of the phonon
spectrum and the prominent peak in the measurement, we present phonon symmetry
representations of related modes and corresponding atomic motions in
Figs. \ref{fig2}(b) and (c). It is seen that the 4-fold branch at
$\sim$170cm $^{-1}$ is formed by the vibrational motions corresponding to the
E$_{2u}$ and E$_{1g}$ modes. Among these only the E$_{1g}$ mode is Raman-active.
However, in backscattering experiments on a surface perpendicular to the c-axis,
the E$_{1g}$ mode is forbidden and is not observed in our experiments.
Furthermore, the decomposition of the six-fold phonon branch at
$\sim$250 cm$^{-1}$ (at $\Gamma$ point) can be described as
$\Gamma=2E_{1u}+2E_{2g}^{1}+B_{1u}+A_{1g}$. Among these the $E_{2g}$ and
$A_{1g}$ modes are Raman-active and are degenerate in energy. As a result of
this degeneracy, only one Raman peak can be observed in our measurements at 250
cm$^{-1}$.

Next, we pay attention to the emergence of the second Raman peak in few and
single layer WSe$_2$ that appears $\sim$11 cm$^{-1}$ above the main peak.
Though, from bulk to single layer WSe$_2$, there is no visible change in the
calculated phonon dispersion (Figs. \ref{fig2}(b,d,e), our measurements show
that a new peak develops (Fig. \ref{fig2}(a)) and this new peak appears to be
very sensitive to the number of layers. For single layer WSe$_2$, decomposition
of the vibration representation is calculated to be
$\Gamma=2E''+2E'+A_{1}'+A_{2}''$. Here, the $E''$, $E'$ and $A_{1}'$ modes
correspond to the $E_{1g}$, $E_{2g}$ and $A_{1g}$ modes for bulk respectively.
Since the $E'$ and $A_{1}$ modes are almost degenerate, only one Raman peak is
expected to be observed in the experiment which contradicts our experimental
results. Before discussing more on the discrepancy, we note that
presence of contaminants at the WSe2/interface and/or directly bonded to the TMD
monolayer might cause small alterations to the Raman spectrum. However, if the
contaminants are directly bonded (chemisorption) to the TMD, one would expect to
observe drastic differences in the Raman spectrum as a result of renormalization
of phonon dispersion. Since the overall monolayer Raman spectrum is similar to
few-layer, we eliminate this possibility. If the contaminants are chemically
interacting locally, this would change the Raman signal from those local regions
(Bloch waves intermix) but would be small comparing to overall Raman signal due
to the geometrical considerations and reduced Raman sensitivity of the locally
interacting region. To explain the emergence of the new peak, we consider
external factors, such as compressive and tensile strain acting on few-layer and
monolayer WSe$_2$ flakes, which is likely to be induced by the interaction with
the substrate.  

To test the possibility of strain effects on WSe$_2$ as the origin for such
splitting of the Raman peak, we calculate the phonon dispersion of monolayer
WSe$_2$ after applying 1$\%$ biaxial compressive and tensile strain. As seen
from Fig.~\ref{fig3}(a), such biaxial deformation only results in a collective
softening/hardening of the vibrational modes and does not lift the degeneracy of
the $E_{2g}$ and $A_{1g}$ modes ($E'$ and $A_{1}'$ in single layer) implying
that these modes remain degenerate as long as the hexagonal symmetry of the
monolayer WSe$_2$ is retained. Next, we apply
uniaxial strain that is likely to occur along with the biaxial strain on
monolayer WSe$_2$ exfoliated on Si/SiO$_2$ surface. For hexagonally
ordered crystal structures uniaxial strain can be applied in two main
directions: armchair and zigzag (Fig.~\ref{fig3}(b)). Interestingly, in the
presence of uniaxial strain, degeneracy of the branches forming the most
prominent Raman peak are broken. While most of the modes between 200-280
cm$^{-1}$ loose their symmetry that determines the Raman activity, our symmetry
analysis shows that two of the branches that correspond to the $E'$ (lower) and
$A_{1}'$ (upper) modes are still Raman active. For compressive (tensile)
uniaxial strain along the armchair direction the difference between the
$E'$ and $A_{1}'$ modes is 16 (18) cm$^{-1}$. Similarly, for compressive
(tensile) strain along the zigzag direction, splitting is calculated to be 13
(17) cm$^{-1}$. Here the $E'$-$A_{1}'$ splitting induced by uniaxial
strain is in good agreement with our experimental data. It is worth to note that
independent from the direction of both tensile and compressive strains both have
the same splitting effect on the Raman peaks. 

\section{Discussions and Conclusions}

According to our phonon dispersion calculations performed on
bulk WSe$_2$, the out-of-plane (A$_{1g}$) and in-plane (E$_{2g}$) modes are
degenerate in energy, consistent with the experimental measurements taken on
bulk WSe$_2$. Experimentally, as the dimension is lowered from 3D to 2D (bulk to
monolayers), Raman mode located at around $\sim$~250 cm$^{-1}$ splits in two
peaks at $\sim$~250 cm$^{-1}$ and 261 cm$^{-1}$. To provide an explanation for
the broken degeneracy, we recalculate the phonon dispersion of monolayer WSe$_2$
under uniaxial and biaxial strain/stress. Our theoretical results imply that
the degeneracy of these two modes is lifted only if the crystal symmetry is
broken, i.e. in the presence of uniaxial strain. When the monolayer WSe$_2$ is
under very little unaxial strain or stress, our theoretical calculations show a
remarkable match with the experimental data. Considering above arguments, we
next argue about the possible origin of uniaxial strain on thin WSe$_2$ flakes.
Since our SiO$_2$ substrates typically display 4-8 \AA~surface roughness
(Fig.~\ref{fig1}(d-e)) that is of the same order
of a single unit cell thickness, even perfect SiO$_2$/WSe$_2$ interface is
likely to induce a mild strain to the WSe$_2$ monolayer. The presence of the
surface roughness on SiO$_2$ can be observed  clearly when compared to sapphire
which yields only 1-2 \AA~surface roughness. Typically, these rough features are
asymmetric in shape and therefore strain induced on the few-layer WSe$_2$ is
expected to have both biaxial and uniaxial component. However, since the biaxial
strain does not lift the degeneracy only the uniaxial component of the total
strain results in Raman peak splitting. Moreover, in the presence of residual
contaminants at the WSe$_2$/SiO$_2$ interface and/or residues deposited after
the exfoliation step, the effect of strain is likely to be amplified. From this
perspective, monolayers are most affected by the strain effects while thick
flakes remain mostly unaltered. Another scenario might be associated with the
unintentional strain induced during the exfoliation process. In such case, the
WSe$_2$ monolayers are deposited on the SiO$_2$ substrates during the
exfoliation and is not necessarily related to the surface residue. However, we
note that our results were confirmed on 50 different WSe$_2$ monolayers where
the same results have been found making this scenario. Another indirect
confirmation of presence of uniaxial strain comes from the changes in the PL
peak position. We also note that the band gap at the K-K symmetry point is
expected to be independent from
the dimensionality. On the contrary, PL measurements show that the K-K gap
increases abruptly for the monolayer, likely due to the presence of small
uniaxial strain (Fig. ~\ref{fig1}(g)) as is confirmed by our DFT calculations.
We also point out that in the presence of large density of defects, the
degeneracy can be lifted. However, similar measurements taken on more than
fifty independently prepared samples makes this case unlikely.

In conclusion, phonon dispersion and the electronic properties
of bulk to monolayer WSe$_2$ have been studied both experimentally and
theoretically. Unlike the other members of the TMDs, the $E_{2g}$ and $A_{1g}$
modes are degenerate in bulk WSe$_2$ and the degeneracy is lifted for as the
dimension is lowered. On the contrary to our experimental results, calculated
phonon dispersion show that these modes remain degenerate independent from the
dimensionality. However, theoretically the degeneracy is only lifted when the
crystal symmetry is broken, i.e. in the presence of unaxial strain which might
be induced by the interaction with the substrate, residues, and/or exfoliation
process. These results provide deeper understanding in the vibrational
properties of TMDs, especially on a material with unique phonon dispersion.

\section{Acknowledgements}

This work was supported by the Flemish Science Foundation (FWO-Vl) and the 
Methusalem programme of the Flemish government. Computational resources were
partially provided by TUBITAK ULAKBIM, High Performance and Grid Computing
Center (TR-Grid e-Infrastructure). H. S. is supported by a FWO Pegasus Marie
Curie Long Fellowship.

\end{document}